\documentclass[a4paper,11pt]{article}
\pdfoutput=1
\usepackage{jheppub}
\pdfoutput=1

\usepackage{epsf}
\usepackage{epsfig}
\usepackage{subfig}
\usepackage{mathtools}
\usepackage{hhline}
\usepackage{float}
\usepackage{multirow}
\usepackage{nicefrac}
\usepackage{epstopdf}
\usepackage{slashed}
\usepackage{xcolor}
\usepackage{url}
\usepackage{braket}
\usepackage{subdepth}
\usepackage{booktabs,siunitx,dcolumn}

\graphicspath{ {figures/} }

\newcommand{\be}{\begin{eqnarray}}
\newcommand{\ee}{\end{eqnarray}}

\newcommand{\ba}{\begin{array}}
\newcommand{\ea}{\end{array}}
\newcommand{\bee}{\begin{equation}\ba{c}}
\newcommand{\eee}{\ea\end{equation}}

\newcommand{\bi}{\begin{itemize}}
\newcommand{\ei}{\end{itemize}}

\allowdisplaybreaks

\def\be{\begin{equation}}
\def\ee{\end{equation}}
\def\bea{\begin{eqnarray}}
\def\eea{\end{eqnarray}}

\def\bbuildrel#1_#2^#3{\mathrel{\mathop{\kern 0pt#1}\limits_{#2}^{#3}}}
\def\slash#1{\setbox0=\hbox{$#1$}#1\hskip-\wd0\dimen0=5pt\advance
       \dimen0 by-\ht0\advance\dimen0 by\dp0\lower0.5\dimen0\hbox
         to\wd0{\hss\sl/\/\hss}}

\def\be{\begin{equation}}
\def\ee{\end{equation}}
\def\beq{\begin{eqnarray}}
\def\eeq{\end{eqnarray}}

\def\slash#1{#1 \hskip-0.45em /}

\def\DB0{\partial B_0}

\def\Cl2{\mbox{Cl}_2}

\def\slash#1{#1 \hskip-0.45em /}

\usepackage{color}

\definecolor{Brown}{rgb}{0.5,0.25,0}

\title{\boldmath Inclusive $\bar{B}\to X_s \ell^+\ell^-$ at the LHC: \\ theory predictions and new-physics reach}

\author{Tobias Huber$^1$,}
\author{Tobias Hurth$^2$,}
\author{Jack Jenkins$^{1}$,}
\author{Enrico Lunghi$^3$,}
\author{Qin Qin$^{4}$,}
\author{K.~Keri Vos$^{5,6}$}
\affiliation{
$^1$Theoretische Physik 1, Center for Particle Physics Siegen (CPPS), Universit\"at Siegen, \\ Walter-Flex-Stra{\ss}e 3, D-57068 Siegen, Germany\\
$^2$PRISMA+ Cluster of Excellence and Institute for Physics (THEP), \\ Johannes Gutenberg University, D-55099 Mainz, Germany\\
$^3$Physics Department, Indiana University, Bloomington, IN 47405, USA \\
$^4$School of Physics, Huazhong University of Science and Technology, Wuhan 430074, China \\
$^5$Gravitational Waves and Fundamental Physics (GWFP), Maastricht University, Duboisdomein 30, NL-6229 GT Maastricht, the Netherlands\\
$^6$Nikhef, Science Park 105, NL-1098 XG Amsterdam, the Netherlands
}
\emailAdd{huber@physik.uni-siegen.de}
\emailAdd{tobias.hurth@cern.ch}
\emailAdd{jack.jenkins@uni-siegen.de}
\emailAdd{elunghi@iu.edu}
\emailAdd{qqin@hust.edu.cn}
\emailAdd{k.vos@maastrichtuniversity.nl}

\abstract{We present theoretical predictions for observables in inclusive $\bar{B}\to X_s \ell^+\ell^-$ suitable for measurements at hadron colliders through a sum-over-exclusive approach. At low $q^2$ we calculate the branching ratio and three angular observables. At high $q^2$ we provide the branching ratio and the ratio of the $\bar B \to X_s \ell^+\ell^-$ rate with respect to the inclusive $\bar B \to X_u \ell \bar\nu$ rate with the same phase-space cut. We compare our predictions to the $B$ factory measurements and also to an extraction of the experimental rate through a sum-over-exclusive method using branching ratios of the exclusive $\bar B \to K^{(*)}\mu^+\mu^-$ modes measured at LHCb. We find a consistent picture comparing Standard Model theory and experiment. As such, our analysis does not support a recent claim about a deficit in the inclusive branching ratio in the high-$q^2$ region. Finally, we present current model-independent bounds on new physics and emphasize the potential of complementary analyses of $\bar{B}\to X_s \ell^+\ell^-$ at Belle~II and the LHC.}

\preprint{
\begin{minipage}{3cm}
\small
\flushright
SI-HEP-2024-07\\
MITP-24-039\\
P3H-24-022\\
Nikhef 2024-005

\end{minipage}}

\begin{document}

\maketitle


\section{Introduction}
\label{sec:introduction}

Flavour-changing neutral transitions of heavy quarks remain prime candidates in the search for physics beyond the Standard Model~(SM) of particle physics. In the exclusive $b\to s$ modes, such as $\bar{B}\to K^{(*)} \ell^+\ell^-$, several deviations between SM predictions and experimental measurements have been found. Theoretically, these predictions are challenging due to nonperturbative charm-loop effects, making a crosscheck via the inclusive modes indispensable and important~\cite{Greljo:2022jac, Ciuchini:2022wbq,Alguero:2023jeh,Hurth:2023jwr}.

Recently, a measurement of the inclusive $\bar B \to X_s \ell^+\ell^-$ at LHCb using isospin symmetry has been proposed~\cite{Amhis:2021oik}. Such a measurement at a hadron collider is challenging and has to rely on the semi-inclusive method (sum over exclusive modes).  
In fact, also at Belle II there will first be a semi-inclusive measurement before a fully inclusive measurement using the recoil technique might become feasible. 

In this letter, we present the corresponding theory predictions, adjusting previous predictions for $\bar B \to X_s \ell^+\ell^-$ observables in the low- and also in the high-$q^2$ region\footnote{Throughout the paper, $q^2$ denotes the di-lepton invariant mass squared.} at Belle~II~\cite{Huber:2005ig,Huber:2007vv,Huber:2015sra,Huber:2020vup} to the environment of the LHC. Specifically, in our previous calculations tailored for the Belle II experiment, we have included the logarithmically enhanced effects of a single-photon emission from the final state leptons in the collinear approximation. This collinear radiation does not affect the rate integrated over the full phase space, but should be taken into account when only parts of the phase space are considered (see Ref.~\cite{Huber:2015sra}, and Ref.~\cite{Bigi:2023cbv} for $\bar B \to X_c \ell\bar\nu$). However, for a future LHCb analysis, such an effect should not be included on the theory side as it is already subtracted using the PHOTOS software\footnote{We thank Patrick Owen, Ulrik Egede and Johannes Albrecht for discussions on this point.}, see also the discussions in Ref.~\cite{Bordone:2016gaq,Isidori:2022bzw,Isidori:2020acz} for the exclusive $B\to K$ mode. Therefore, we present here our results without these collinear QED logarithms.

The theoretical description of the $\bar B \to X_s \ell^+\ell^-$ decay includes 
NNLO QCD~\cite{Bobeth:1999mk,Asatryan:2001zw,Asatrian:2001de,Asatryan:2002iy,Ghinculov:2002pe,Asatrian:2002va,Ghinculov:2003qd,Greub:2008cy,deBoer:2017way}, NLO electroweak  corrections \cite{Huber:2005ig,Bobeth:2003at,Huber:2007vv,Huber:2015sra}, and $1/m_b^2$ \cite{Falk:1993dh,Ali:1996bm,Chen:1997dj,Buchalla:1998mt} and $1/m_b^3$ \cite{Bauer:1999kf,Ligeti:2007sn} power-corrections. The five-particle contributions to $\bar B \to X_s \ell^+\ell^-$ are numerically negligible~\cite{Huber:2018gii}.
Long-distance off-shell effects from intermediate charmonium resonances are included through the Kr\"uger-Sehgal (KS) approach~\cite{Kruger:1996cv,Kruger:1996dt} and the $1/m_c^2$ corrections~\cite{Buchalla:1997ky}   (see also the detailed discussions in Refs.~\cite{Ghinculov:2003qd,Huber:2019iqf})\footnote{In this way there is no double counting within the off-shell effects of the resonances. While the KS method is exact when only factorisable contributions are taken into account, this means that the $\bar B \to X_s  c \bar c$ transition can be factorised into the product of $\bar s b$ and $c \bar c$ colour singlet currents, the $1/m_c^2$ corrections represent the nonfactorisable $c \bar c$    long distance effects far from the resonances~\cite{Ghinculov:2003qd,Huber:2019iqf}}. We also mention here that in the low-$q^2$ region  the local $1/m^2_c$ corrections get overwritten by the resolved (non-local) contributions~\cite{Benzke:2017woq,Hurth:2017xzf,Benzke:2020htm} and should be replaced by the latter.

In the high-$q^2$ region, theoretical predictions of the branching ratio are very sensitive to power corrections arising in the employed operator-product expansion (OPE) framework. In addition, four-quark operators enter at order $1/m_b^3$ and higher. In order to reduce the sensitivity to these power corrections, it is interesting to consider the ratio \cite{Ligeti:2007sn}\footnote{In this ratio, the nonperturbative corrections related to the dominant $Q_9$ and $Q_{10}$ operators cancel out.}
\begin{equation}\label{eq:Rzoldef}
    \mathcal{R}(q_0^2) = \int_{q_0^2}^{M_B^2} dq^2 \frac{d\Gamma(\bar B \to X_s \ell^+\ell^-)}{dq^2} \left/ \int_{q_0^2}^{M_B^2}dq^2 \frac{d\Gamma(\bar{B}\to X_u \ell\bar\nu)}{dq^2} \right. \ ,
\end{equation}
where both the $b\to s$ and $b\to u$ decays have the same lower cut $q_0^2$. In the low-$q^2$ region the ratio $\mathcal{R}$ has reduced sensitivity to the hadronic mass cut which disrupts the local OPE~\cite{Lee:2005pwa, Lee:2008xc, Huber:2023qse}. We note that in this ratio, we should also consider the treatment of the log-enhanced QED terms in the $b\to u$ transition. After considering the experimental setup, we should not include such terms\footnote{We thank Florian Bernlochner and Lu Cao for discussions on this point.}. Hence, we present new calculations of this ratio, without log-enhanced QED corrections in both the $b\to u$ and $b\to s$ mode. 

Finally, we study constraints on new physics from $\bar B \to X_s\ell^+\ell^-$ in the high- and low-$q^2$ regions. The new-physics reach is strengthened by the interplay between (semi)-inclusive analyses at $e^+e^-$ and hadron machines, and between inclusive and exclusive modes.

This article is organised as follows. We present our updated input parameters in section~\ref{sec:inputs}, which lead to our phenomenological results which we present in section~\ref{sec:phenoresults}. Here we also discuss the comparison between the direct calculation of the high-$q^2$ branching ratio in the OPE, versus that obtained through $\mathcal{R}$ in \eqref{eq:Rzoldef} multiplied by the experimental results for the $\bar{B}\to X_u\ell\bar\nu$ decay. We compare to measurements from BaBar~\cite{BaBar:2004mjt, BaBar:2013qry} and Belle~\cite{Belle:2005fli}. Moreover, we present a first sum-over-exclusive determination purely from experimental measurements of exclusive modes such as $\bar{B}\to K^{(*)} \mu^+\mu^-$ from LHCb. We also comment on a recent calculation of $\mathcal{R}$, and critically reassess a claim about a deficit in the data within the high-$q^2$ region~\cite{Isidori:2023unk}.  In section~\ref{sec:np}, we present our analysis of the new-physics reach, while we conclude in section~\ref{sec:conclusion}.
Additional numerical results including log-enhanced QED corrections are relegated to appendix~\ref{sec:app:results}.


\section{Input parameters}
\label{sec:inputs}

In the high-$q^2$ region, the uncertainties of both the branching ratio $\cal B$ and the ratio $\mathcal{R}$ are dominated by power corrections to the OPE proportional to $\lambda_i$ and $\rho_i$ as defined in our previous work~\cite{Huber:2019iqf,Huber:2020vup}. The expressions of these observables with the heavy-quark expansion (HQE) parameters left symbolic are
\begin{align}
    {\cal{B}}[>14.4] &=  (3.05  - 5.87~\lambda_2^{\rm eff} + 8.09~\rho_1) \times 10^{-7} \, , \label{eq:sym} \\[0.5em]
    {\cal{B}}[>15]   &=  (2.55  - 5.717~\lambda_2^{\rm eff} + 8.47~\rho_1) \times 10^{-7} \, , \label{eq:sym2} \\[0.9em]
    \mathcal{R}(14.4) &=  (24.90 + 0.04~\lambda_1 + 2.49~\lambda_2^{\rm eff} + 10.72~\rho_1) \times 10^{-4} \, , \label{eq:sym3} \\[0.5em]
    \mathcal{R}(15)   &=  (25.65 + 0.03~\lambda_1 + 3.66~\lambda_2^{\rm eff} + 11.94~\rho_1) \times 10^{-4} \, , \label{eq:sym4}
\end{align}
where 
\begin{equation}
      \lambda_2^{\rm eff} = \lambda_2 - \frac{\rho_2}{m_b} \, .
\end{equation}
Here, the numerical values for $\lambda_1$ and $\lambda_2^{\rm eff}$ have to be inserted in units of GeV$^2$, that of $\rho_1$ in units of GeV$^3$. The effect of $\lambda_2^{\rm eff}$ and in particular $\rho_1$ is pronounced, and the dependence of the branching ratio on the HQE parameters is larger compared to the ratio $\mathcal{R}$~\cite{Ligeti:2007sn}. The dependence of the branching ratios on $\lambda_1$ is marginal and not given separately. In addition, the effect of four-quark operators~\cite{Voloshin:2001xi,Ligeti:2007sn,Gambino:2010jz,Huber:2019iqf} is large, but we do not give their dependence here explicitly.

In this chapter we update the values for the HQE parameters $\lambda_{1,2}$ and $\rho_1$. All remaining input parameters are the same as in table~1 of Ref.~\cite{Huber:2020vup}. The HQE parameters are extracted from the $\bar B \to X_c\ell\bar\nu$ spectrum. We use the results from Ref.~\cite{Bordone:2021oof}, which defines the HQE elements for the physical states. A first analysis using only di-lepton invariant mass moments has also become available \cite{Bernlochner:2022ucr}, which uses the reparametrization invariant basis \cite{Fael:2018vsp} and includes also higher-order HQE elements. In addition, a new analysis appeared including all available moments of the spectrum \cite{Finauri:2023kte}. The latter results are in agreement with the results in Ref.~\cite{Bordone:2021oof}, which we use in our analysis. 

As is by now customary, the HQE parameters are extracted in the kinetic scheme \cite{Bigi:1994ga, Bigi:1996si} at a scale $\mu=1$ GeV. While for the $b$ quark mass, we work in the $1$S scheme as previously, we use the kinetic scheme for the HQE power corrections. To do so, it requires transforming from the pole scheme expression in Eq.~\eqref{eq:sym} to the kinetic scheme using the perturbative corrections at the appropriate order (see e.g. Refs.~\cite{Gambino:2004qm,Alberti:2014yda} and Ref.~\cite{Fael:2020njb} for the recent update to ${\cal O}(\alpha_s^3)$). Since we work at NNLO, we apply only the ${\cal O}(\alpha_s^2)$ perturbative corrections. As previously, we convert the HQE parameters in the kinetic scheme to their pole scheme counterparts. Using the $\rho_D^3(\mu = 1 \; \rm{GeV})$ and $\mu_\pi^2(\mu = 1 \; \rm{GeV})$ values from \cite{Bordone:2021oof}, we then find
\begin{equation}
    \rho_1 =\rho_D^3(0) = (0.080 \pm 0.031)\; \rm{GeV}^3 \ , \quad -\lambda_1 = \mu_\pi^2(0)= (0.314 \pm 0.056)\; \rm{GeV}^2 \ ,
\end{equation}
where contrary to the analysis in Ref.~\cite{Huber:2019iqf}, we do not add an additional uncertainty due to missing higher orders. 
In addition, we have 
\begin{equation}
    \lambda_2^{\rm eff} = 
    \frac{\mu_G^2}{3} - \frac{\rho_{LS}^3}{m_b} = (0.111 \pm 0.018)\; {\rm GeV}^2 \,
\end{equation}
where $\mu_G^2$ and $\rho_{LS}^3$ do not receive any perturbative corrections when changing from the pole to the kinetic scheme or vice versa. To facilitate the comparison with our previous analysis~\cite{Huber:2020vup}, we quote the values of the HQE parameters used in~\cite{Huber:2019iqf,Huber:2020vup} as $\rho_1 = (0.038 \pm 0.070)\; \rm{GeV}^3$, $\lambda_1 = (-0.267 \pm 0.090)\; \rm{GeV}^2$, and $\lambda_2^{\rm eff} = (0.130 \pm 0.021)\; {\rm GeV}^2$.

Besides, also four-quark operators (weak annihilation matrix elements) enter. Following the definitions in Refs.~\cite{Huber:2019iqf,Huber:2020vup}, we stress again that the ratio $\mathcal{R}$ is rather insensitive to these corrections and only depends on $f_s (f_{\rm NV})$ and $\delta f$, which measures the SU(3) breaking between the strange and $u$ four-quark operators. On the other hand, the branching ratio depends on both $f_{\rm NV}$ and $f_{\rm V}$, the non-valence and valence weak annihilation contributions, respectively, which have a large uncertainty. We do not update these values here and use those given in Refs.~\cite{Huber:2019iqf,Huber:2020vup}.

Finally, we note that in \cite{Huber:2019iqf,Huber:2020vup} the branching ratio (and other angular observables) were obtained by normalizing to the ratio 
\begin{equation}
    C \equiv \left|\frac{V_{ub}}{V_{cb}}\right|^2 \frac{\Gamma(\bar{B}\to X_c e\bar{\nu})}{\Gamma(\bar{B}\to X_u e\bar{\nu})} 
\end{equation}
to suppress the weak annihilation contributions which enter both the neutral-current $\bar B \to X_s \ell^+\ell^-$ and the charged-current $\bar B \to X_u \ell\bar\nu$ modes. Here we treat $C$ as an input, which we take as in Ref.~\cite{Huber:2020vup} although in principle this parameter will have shifted due to the updated power correction parameters (and higher-order $\alpha_s^3$ corrections now available \cite{Chen:2023dsi,Fael:2023tcv}). In a future analysis, it may be interesting to reconsider this normalization channel.


\section{Phenomenological results}
\label{sec:phenoresults}

\subsection{Standard Model predictions for LHCb} \label{sec:smpred}
\begin{table}
	\begin{center}
		\begin{tabular}{|cc|cc|cc|cc|}
			\hline 
		\multicolumn{2}{|c|}{\rule{0pt}{12.8pt}$q^2~\rm{range} \;[{\rm GeV}^2]$}	& \multicolumn{2}{c|}{$[1,6]$} & \multicolumn{2}{c|}{$[1,3.5]$} & \multicolumn{2}{c|}{$[3.5,6]$} \\[0.1em]
		  \hline
        \multicolumn{2}{|c|}{\rule{0pt}{12.8pt}${\cal B}~[10^{-7}]$} & \multicolumn{2}{c|}{$16.87 \pm 1.25$} & \multicolumn{2}{c|}{$9.17\pm 0.61$} & \multicolumn{2}{c|}{$7.70\pm0.65 $} \\[0.3em]
		\multicolumn{2}{|c|}{${\cal H}_T~[10^{-7}]$} & \multicolumn{2}{c|}{$3.14\pm 0.25$} & \multicolumn{2}{c|}{$1.49\pm 0.09$} & \multicolumn{2}{c|}{$1.65\pm 0.17$} \\[0.3em]
		\multicolumn{2}{|c|}{${\cal H}_L~[10^{-7}]$} & \multicolumn{2}{c|}{$13.65\pm 1.00$} & \multicolumn{2}{c|}{$7.63\pm 0.54$} & \multicolumn{2}{c|}{$6.02\pm 0.49$} \\[0.3em]
		\multicolumn{2}{|c|}{${\cal H}_A~[10^{-7}]$} & \multicolumn{2}{c|}{$-0.27\pm 0.21$} & \multicolumn{2}{c|}{$-1.08\pm 0.08$} & \multicolumn{2}{c|}{$0.81\pm 0.16$} \\[0.1em]
			\hline \hline
		\multicolumn{2}{|c|}{\rule{0pt}{12.8pt}$q^2~\rm{range} \;[{\rm GeV}^2]$} & \multicolumn{3}{c|}{\quad \, $>14.4$ \quad\quad\, } & \multicolumn{3}{c|}{$>15 $}\\[0.1em] \hline
		\multicolumn{2}{|c|}{\hspace*{16pt} ${\cal B}~[10^{-7}]$} & \multicolumn{3}{c|}{\quad\,\rule{0pt}{12.8pt}$3.04\pm 0.69$\quad\quad\,} & \multicolumn{3}{c|}{\rule{0pt}{12.8pt}$2.59\pm 0.68$}\\[0.3em]
		\multicolumn{2}{|c|}{$\mathcal{R}(q^2_0)~[10^{-4}]$} & \multicolumn{3}{c|}{\quad\,$26.02 \pm 1.76$\quad\quad\,} & \multicolumn{3}{c|}{$ 27.00 \pm 1.94$}\\[0.1em] \hline
			\end{tabular}
		\caption{Phenomenological results without logarithmically enhanced electromagnetic effects. The slight changes compared to \cite{Huber:2020vup} are due to the change in the input parameters.\label{tab:phenoresults}}
	\end{center}
\end{table}

As outlined in the introduction, we calculate the SM predictions for inclusive $\bar B \to X_s \ell^+\ell^-$ without logarithmically enhanced electromagnetic corrections, to match the LHCb conditions. We give the branching ratio and $\mathcal{R}(q_0^2)$ in the high-$q^2$ region above $q_0^2 = 14.4$ and 15.0 GeV$^2$. In the low-$q^2$ region, from $1~{\rm GeV}^2 < q^2 < 6$ GeV$^2$, we present the branching ratio as well as the angular coefficients ${\cal H}_T, {\cal H}_L$ and ${\cal H}_A$ defined in Refs.~\cite{Huber:2015sra,Huber:2020vup}. These numbers (except for the $q^2>15$ GeV$^2$ observables) were already given in Table~5 of Ref.~\cite{Huber:2020vup}, albeit without uncertainties. In Table~\ref{tab:phenoresults}, we give these observables and those for $q^2>15$ GeV$^2$ for our updated input parameters and including uncertainties. The change of the HQE parameters alone amounts to a 17\% increase of $\mathcal{B}[>14.4]$, while $\mathcal{R}(14.4)$ changes at the same time by only $+1.6$\%~\footnote{Note that the number for $\mathcal{B}[>14.4]$ in Table~5 of~\cite{Huber:2020vup} still contains log-enhanced QED corrections in the $b\to u \ell \nu$ normalization channel, which have to be removed before the bare change in input parameters becomes accessible.}, which is expected from eq.~\eqref{eq:sym3}. In both cases we observe a slight decrease of the uncertainty. In the low-$q^2$ region, the dominant contributions to the uncertainties stem from the resolved contribution~\cite{Benzke:2017woq,Hurth:2017xzf,Benzke:2020htm} ($5\%$ of the total) and the renormalisation scale (around $3\%$).

In the high-$q^2$ region, we also provide below the numbers for the integrated branching ratio for two different values of the lower $q^2$-cutoff and with individual uncertainties. The total uncertainty is obtained by combining the individual ones in quadrature.
\begin{align}
     \mathcal{B}[>14.4] &= (3.04 \pm 0.25_{\rm scale} \pm 0.03_{m_t} \pm 0.04_{C,m_c} \pm 0.22_{m_b} \pm 0.005_{\alpha_s} \pm 0.003_{\rm CKM} \nonumber \\
     & \hspace*{38pt}\pm 0.05_{\rm BR_{\rm sl}} \pm 0.25_{\rho_1} \pm 0.11_{\lambda_2} \pm 0.54_{f_{u,s}}) \times  10^{-7} \nonumber \\
     &= (3.04 \pm 0.69) \times 10^{-7}\;,  \label{eq:branc2}    \\[0.6em]
          \mathcal{B}[>15] &= (2.59 \pm 0.21_{\rm scale} \pm 0.03_{m_t} \pm 0.05_{C,m_c} \pm 0.19_{m_b} \pm 0.004_{\alpha_s} \pm 0.002_{\rm CKM} \nonumber \\
     & \hspace*{38pt}\pm 0.04_{\rm BR_{\rm sl}} \pm 0.26_{\rho_1} \pm 0.10_{\lambda_2} \pm 0.54_{f_{u,s}}) \times  10^{-7} \nonumber \\
     &= (2.59 \pm 0.68) \times 10^{-7} \label{eq:branc}\; .
\end{align}
We note that the uncertainty is still dominated by the uncertainty on the higher-order HQE parameters and four-quark operators, but could be already reduced due to the improved input parameters. Moreover, we observe that the branching ratio drops rather steeply as the lower $q^2$-cutoff is increased. The branching ratio is $\sim 15$\% smaller when integrating from $15~\rm{GeV}^2$ compared to $14.4~\rm{GeV}^2$. Hence both, from the point of view of statistics, and on theoretical grounds a lower cutoff $q_0^2 = 14.4~\rm{GeV}^2$ is favourable. In the high-$q^2$ region, one encounters a breakdown of the heavy-mass expansion at the endpoint of the spectrum. However, for an {\emph{integrated}} spectrum an effective expansion in inverse powers of $m_b^{\rm{eff}} = m_b (1 - \sqrt{q_0^2/m_b^2})$ rather than $m_b$ is found~\cite{Buchalla:1998mt,Neubert:2000ch,Bauer:2001rc,Ghinculov:2003qd}, which is expected to behave better the lower $q_0^2$ is.

Alternatively, we can calculate the ratio $\mathcal{R}(q_0^2)$ defined in Eq.~\eqref{eq:Rzoldef}, which has a reduced sensitivity to the power corrections. Excluding logarithmically enhanced electromagnetic corrections in both the $b\to s$ and $b\to u$ transition,
we obtain
\begin{align}
\allowdisplaybreaks
    \mathcal{R}(14.4) =&\ (26.02 \pm 0.42_{\rm scale} \pm 0.30_{m_t} \pm 0.11_{C,m_c} \pm 0.10_{m_b} \pm 0.12_{\alpha_s} \pm 1.12_{\rm CKM}  \nonumber \\
     & \hspace*{33pt} \pm 0.33_{\rho_1} \pm 0.05_{\lambda_2} \pm 1.20_{f_{u,s}}) \times  10^{-4}  \nonumber \\
     =& \ (26.02 \pm 1.76) \times 10^{-4} \,, \label{eq:Rzol1} \\[1.8em]
    \mathcal{R}(15)   =&\ (27.00 \pm 0.25_{\rm scale} \pm 0.30_{m_t} \pm 0.11_{C,m_c} \pm 0.17_{m_b} \pm 0.15_{\alpha_s} \pm 1.16_{\rm CKM} \nonumber \\
     & \hspace*{33pt}\pm 0.37_{\rho_1} \pm 0.07_{\lambda_2} \pm 1.43_{f_{u,s}}) \times  10^{-4}  \nonumber \\
     = &\ (27.00 \pm 1.94) \times 10^{-4} \, . \label{eq:Rzol2}
\end{align}

\begin{figure}
\begin{center}
\includegraphics[width=0.55\linewidth]{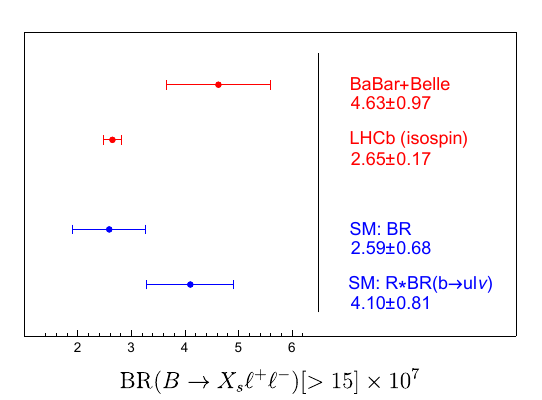}
\caption{Comparison between the high-$q^2$ branching ratio obtained from BaBar and Belle in \eqref{eq:babarbelle} and a sum over exclusive states using LHCb data. In addition, we show the direct SM theory prediction in \eqref{eq:branc} and that obtained by multiplying the ratio ${\cal R}$ by the experimental $\bar{B}\to X_u \ell \bar\nu$ rate from~(\ref{eq:TheoryViaR15}).}
\label{fig:semicomp}
\end{center}
\end{figure}

We can now use the $\cal{R}$ ratio combined with the latest Belle measurement of the inclusive branching ratio $\mathcal{B}(\bar B \to X_u \ell\bar\nu)$  \cite{Belle:2021ymg} to obtain the inclusive branching ratio $\mathcal{B}(\bar B \to X_s \ell^+\ell^-)$~\cite{Isidori:2023unk}. From Ref.~\cite{Belle:2021ymg}\footnote{This result was obtained using the $q^2$ differential data provided by the Belle collaboration in https://doi.org/10.17182/hepdata.131599}, we find
\begin{align} 
\mathcal{B}( \bar{B} \to X_u \ell \bar\nu)[>14.4]_{\rm exp}  = (1.76 \pm 0.32)\times 10^{-4}\, , \label{eq:BXuexp}  \\
 \mathcal{B}( \bar{B} \to X_u \ell \bar\nu)[>15]_{\rm exp}
 = (1.52 \pm 0.28)\times 10^{-4}\, ,
\end{align} 
where the latter is in agreement with the value quoted in Ref.~\cite{Isidori:2023unk}. 
Using now our theoretical prediction for the ratio $\cal{R}$, we find
\begin{align}
\mathcal{B}[>14.4]_{\rm SM, {\cal R}} &= {\cal R}(14.4) \times \mathcal{B}( B \to X_u \ell \bar\nu)[>14.4]_{\rm exp} \nonumber \\
{}&=  (4.58 \pm 0.89) \times 10^{-7} \,. 
\label{eq:TheoryViaR14.4}\\
\mathcal{B}[>15]_{\rm SM, {\cal R}} {}&= {\cal R}(15) \times \mathcal{B}( B \to X_u \ell \bar\nu)[>15]_{\rm exp} \nonumber \\
{}&=  (4.10 \pm 0.81 ) \times 10^{-7} \,.
\label{eq:TheoryViaR15}
\end{align}
These determinations are compatible with our direct calculations in Eqs.~(\ref{eq:branc2}) and~(\ref{eq:branc}).  The comparison between the two values is given in Figure~\ref{fig:semicomp}. It will be interesting to see how the central value of $\mathcal{B}(\bar{B}\to X_u\ell \bar{\nu})$ will develop in the future with more statistics. Also the power corrections in the branching ratios of Eqs.~(\ref{eq:branc2}) and~(\ref{eq:branc}) will be scrutinized in the future. 

Finally, let us briefly comment on the prediction of the ratio ${\cal R}$ in Ref.~\cite{Isidori:2023unk}. The authors find a larger central value and also a larger uncertainty compared to our prediction. To further analyse this issue, we divide our central value $\mathcal{R}(15) = 0.0027$ by the ratio $|V_{tb} V_{ts}^\ast/V_{ub}|^2 = 123.5$ of CKM elements (see Table~1 of Ref.~\cite{Huber:2020vup}), and obtain a value of $2.19 \times 10^{-5}$, which turns out to be $13\%$ smaller than the central value $2.51 \times 10^{-5}$ 
obtained from Eq.~(22) in~Ref.~\cite{Isidori:2023unk}. The difference can be traced back to the fact that we include the two-loop ${\cal O}(\alpha_s^2)$-corrections induced by the mixing of current-current operators into $Q_7$ and $Q_9$~\cite{Ghinculov:2003qd,Greub:2008cy,deBoer:2017way,Asatrian:2019kbk}, as well as the factorisable $c\bar c$ long distance off-shell effects via the KS-approach~\cite{Kruger:1996cv,Kruger:1996dt}. Both contributions seem to be absent in the analysis of Ref.~\cite{Isidori:2023unk} -- which largely relies on the implementation in Ref.~\cite{Ligeti:2007sn} -- and result in a correction in the same direction.

Adding the effect of the $Q_{1,2}-Q_{7,9}$ interference at order ${\cal O}(\alpha_s^2)$   shifts $\mathcal{R}(15)$ by $\sim - 9\%$, which reveals that this is the main source of discrepancy with our result. The remaining difference can be attributed to the factorisable $c\bar c$ long distance off-shell effects via the KS-approach; adding the latter contribution leads to an additional shift of $\mathcal{R}(15)$ by $\sim - 4\%$. When adding these two contributions to the result in Eq.~(22) of Ref.~\cite{Isidori:2023unk}, we still find a small difference at the one-percent level with our result. We conjecture that the tiny difference originates from different input parameters and the fact that the authors of Ref.~\cite{Isidori:2023unk} implement the full NLO electroweak matching at the high scale~\cite{Bobeth:2013tba}, while our code contains this contribution only in the large-$m_t$ limit (see Ref.~\cite{Buchalla:1997kz} and references therein). 

As stated earlier, it is unarguable that the two-loop QCD matrix elements of the $Q_{1,2}-Q_{7,9}$ interference~\cite{Ghinculov:2003qd} must be included, as well as the long-distance off-shell effects from the intermediate charmonium resonances, the factorisable ones via the KS approach~\cite{Kruger:1996cv,Kruger:1996dt} and the nonfactorisable ones via $1/m_c^2$ corrections~\cite{Buchalla:1997ky}   (see also discussions in Refs.~\cite{Ghinculov:2003qd,Huber:2019iqf}).

\subsection{Sum-over-exclusive from LHCb}

Although inclusive $\bar{B} \to X_s \ell^+ \ell^-$ observables have only been measured at the B factories~\cite{Belle:2005fli, Belle:2014owz, BaBar:2004mjt, BaBar:2013qry}, in the high-$q^2$ region it is interesting to consider a semi-inclusive determination from a sum over exclusive $\bar{B}\to K(n\pi)$ modes measured at LHCb \cite{LHCb:2014cxe,LHCb:2016ykl,LHCb:2014osj}. For $q^2>15\, \textrm{GeV}^2$, the invariant mass of the $X_s$ system is constrained by kinematics to $M_{X_s}<1.41 \, \textrm{GeV}$. Therefore the inclusive rate is expected to be saturated by a limited number of exclusive modes, and dominated by the $K$ and $K^*(980)$, with a modest contribution from the broad $S$-wave $K\pi$ modes. Higher-mass resonances, like the $K^*(1410)$ and $K^*(1430)$ states, are kinematically suppressed in the high-$q^2$ region, although their tail may have some effect in the $K\pi$ and $K\pi\pi$ modes. For the purposes of a semi-inclusive determination it would be sufficient to measure each mode (without partial wave expansion) integrated over the full hadronic phase space to account for such effects.

The $\bar{B} \to K^{(*)}\mu\mu$ branching ratios are available for both charged and neutral $B$ mesons~\cite{LHCb:2014cxe,LHCb:2016ykl}. The weighted isospin averages of the two pairs of measurements\footnote{In the following we denote the two charged mesons $B^\pm$ by $B^+$ and the neutral $B^0, \bar{B}^0$ mesons by $B^0$. We denote the isospin and CP averaged (time-integrated) ensemble as $\bar{B}$.}\footnote{The systematic uncertainties could be reduced by updating the input for the $B\to J/\psi K$ normalization to its current world-average \cite{10.1093/ptep/ptac097}.}
\begin{align}
\mathcal{B}(B^+\to K^+ \mu \mu)[>15] &= (0.85 \pm 0.05) \times 10^{-7} \, , \label{eq:BtoK1}\\
\mathcal{B}(B^0 \to K^0 \mu \mu)[>15] &= (0.67 \pm 0.12) \times 10^{-7} \label{eq:BtoK2}
\end{align}
and
\begin{align}
\mathcal{B}(B^+ \to K^{*+}\mu\mu)[>15] &= (1.58 \pm 0.32) \times 10^{-7} \, , \label{eq:BtoKs1} \\
\mathcal{B}(B^0 \to K^{*0}\mu\mu)[>15]&=(1.74 \pm 0.14)\times 10^{-7} \label{eq:BtoKs2}
\end{align}
are given by
\begin{align}
    \mathcal{B}(\bar{B} \to K \mu \mu)[>15] &= (0.82 \pm 0.05) \times 10^{-7} \, , \label{eq:Kavg} \\ 
    \mathcal{B}(\bar{B} \to K^* \mu \mu)[>15] &= (1.72 \pm 0.13) \times 10^{-7} \, . \label{eq:Kstaravg}
\end{align}
The averages above do not account for experimental correlation, but are dominated by the more precise results in Eqs.~\eqref{eq:BtoK1} and~\eqref{eq:BtoKs2}. Therefore even severe correlations between Eqs.~(\ref{eq:BtoK1},~\ref{eq:BtoK2}) or Eqs.~(\ref{eq:BtoKs1},~\ref{eq:BtoKs2}) would only marginally increase the uncertainties quoted in Eqs.~\eqref{eq:Kavg} and~\eqref{eq:Kstaravg}. The sum of the $K$ and $K^*$ modes is
\begin{equation}\label{eq:av}
    \mathcal{B}(\bar{B} \to K^{(*)} \mu \mu)[>15] = (2.54 \pm 0.14) \times 10^{-7} \, ,
\end{equation}
where we have added the uncertainties in Eqs.~\eqref{eq:Kavg} and~\eqref{eq:Kstaravg} in quadrature, a conservative accounting of systematics common to $K$ and $K^*$ measurements.

The LHCb collaboration also measured the $S$-wave fraction in $B^0 \to K^+ \pi^- \ell^+ \ell^-$ decay~\cite{LHCb:2016ykl}. The hadronic phase space $0.64 \, \textrm{GeV} < M_{K\pi} < 1.20 \, \textrm{GeV}$ used in this analysis includes nearly all of the phase space in the high-$q^2$ region. Due to the large uncertainties at high-$q^2$ and coarse binning, we do not attempt to extrapolate to the remaining phase space here, nor do we assign an additional uncertainty to this effect. The $B^0 \to K^0 \pi^0\mu\mu$ and $B^+ \to (K^+ \pi^0, K^0 \pi^+)\mu\mu$ modes with neutrals in the final state can be estimated using isospin relations (see for instance Eq.~(48) of Ref.~\cite{Buchalla:1998mt}). These relations are summarized in the following expression for the isospin ($B^+, B^0$) average of the sum of $K\pi$ modes,
\begin{align}
    \mathcal{B}(\bar{B} \to (K \pi)_J \, \ell^+ \ell^-) = \mathcal{B}(\bar{B} \to (K^+ \pi^-)_J \,  \ell^+ \ell^-) &\times \begin{cases}
        \frac{3}{2} & J=0 \\
        1 & J=1
    \end{cases} \label{eq:RKpi}
\end{align}
where $J$ indicates the total angular momentum of the system of considered particles. From the measured $S$-wave fraction in \cite{LHCb:2016ykl}, normalized using the $B^+ \to K^{*0}\mu \mu$ rate in Eq.~\eqref{eq:BtoK2} and multiplied by the factor of $3/2$ in Eq.~\eqref{eq:RKpi}, we obtain
\begin{align}
    \mathcal{B}(\bar{B} \to (K\pi)_S \mu \mu)[>15] &= (0.05 \pm 0.09) \times 10^{-7} \, . \label{eq:Skpi}
    \end{align}
Although consistent with zero, Eq.~\eqref{eq:Skpi} has a smaller uncertainty than a theoretical estimate using chiral perturbation theory $(0.58 \pm 0.25) \times 10^{-7}$ of the same quantity~\cite{Isidori:2023unk}. 

For the $K\pi\pi$ mode, only the $B^+ \to K^+ \pi^+ \pi^- \ell^+ \ell^-$ branching ratio has been observed~\cite{LHCb:2014osj}. The isospin average rate decomposed in $\pi\pi$ partial waves is
\begin{align}
    \mathcal{B}(\bar{B} \to K (\pi \pi)_S\, \ell^+ \ell^-) = \mathcal{B}(B^+ \to K^+ (\pi^+ \pi^-)_S \, \ell^+ \ell^-) \times \frac{3}{2} \, . \label{eq:RKpipi}
\end{align}
Here we assume that the $K\pi\pi$ mode is dominated by $\pi\pi$ in $S$-wave. This is a reasonable assumption given that $m_{\pi\pi}<0.91\,\textrm{GeV}$ in the high-$q^2$ region. 
We obtain\footnote{To be precise, this number corresponds to the branching ratio of the $K\pi\pi$ mode for $q^2>14.18 \, \textrm{GeV}^2$, an upper limit on $q^2>15 \, \textrm{GeV}^2$.}
\begin{align}\label{eq:2picon}
    \mathcal{B}(B \to K\pi\pi \mu \mu)[>15] &= (0.06 \pm 0.04) \times 10^{-7} \, .
\end{align}

In principle, also even higher multiplicity modes with $B \to K(>2\pi) \mu \mu$ could have an effect. 
As such modes are expected to be further suppressed, we estimate
\begin{align}
\mathcal{B}(B \to K(>2\pi) \mu \mu)[>15] = (0.00 \pm 0.04) \times 10^{-7} \, , \label{eq:K3pi}
\end{align}
where the uncertainty is chosen to be the same as in Eq.~\eqref{eq:2picon}. 
Combining Eqs.~\eqref{eq:Skpi}, \eqref{eq:2picon} and~\eqref{eq:K3pi} we obtain for the non-$K^{(*)}$ contributions
\begin{align}
    \mathcal{B}(B \to K(n\pi)\mu\mu)[>15] &= (0.11 \pm 0.10) \times 10^{-7} \ . \label{eq:Knpi}
\end{align}
Combining this result with the $K^{(*)}$ modes in Eq.~\eqref{eq:av}, we finally obtain\footnote{This number was obtained in collaboration with the authors of \cite{Isidori:2023unk}.}
\begin{align}
      \mathcal{B}[>15]_{\rm LHCb \, (isospin)}  = (2.65 \pm 0.17) \times 10^{-7}\,.
    \label{eq:ExpSemiBR2}
\end{align}
Comparing the number above to Eq.~\eqref{eq:av}, it is remarkable that the data from LHCb indicate that the $K^{(*)}$ modes saturate $\sim 95\%$ of the inclusive rate in the high-$q^2$ region. It will be interesting to scrutinize this empirical observation with more precise experimental determinations of the remaining contributions in Eqs.~\eqref{eq:Knpi}. 

It is interesting to compare the determination of the inclusive rate in Eq.~\eqref{eq:ExpSemiBR2} to the inclusive measurements at Belle~\cite{Belle:2005fli, Belle:2014owz} and BaBar~\cite{BaBar:2004mjt, BaBar:2013qry}, for $q^2>14.4 \, \textrm{GeV}^2$ and $q^2>14.2 \, \textrm{GeV}^2$ respectively. 
Due to the different bins and the different treatment of the QED effects, the B factory results cannot be directly compared to Eq.~\eqref{eq:ExpSemiBR2}. In order to compare, we use the ratios of our theoretical predictions of the inclusive rate for the different bins and QED treatments (see Tables~\ref{tab:phenoresults} and~\ref{tab:phenoresultsQED})\footnote{We do not provide a prediction $\mathcal{B}[>14.2]$ for the B factories in the Tables.} to obtain the rescaling factors
\begin{align}
    \frac{\mathcal{B}[>14.4]_\textrm{with QED}} { \mathcal{B}[>14.2]_\textrm{with QED}} &= 0.96  \, ,\\
    \frac{ \mathcal{B}[>15]_\textrm{no QED} } { \mathcal{B}[>14.4]_\textrm{with QED} } &= 0.97 \, , \label{eq:eps}
\end{align}
where the former is used to interpolate the BaBar result to $q^2>14.4$, and the latter is used to compare the resulting B factory average at $q^2>14.4$ to Eq.~\eqref{eq:ExpSemiBR2} as follows (see also the more detailed discussion in Section~\ref{sec:averageexptheo}). Note that the correction in Eq.~\eqref{eq:eps} is marginal due to a partial cancellation of phase space enhancement and QED bin migration. We obtain
\begin{equation}
    \mathcal{B}[>15]_{\rm BaBar/Belle} = (4.63\pm 0.97) \times 10^{-7} \,.
\label{eq:babarbelle}
\end{equation}

In Fig.~\ref{fig:semicomp}, we compare Eqs.~\eqref{eq:ExpSemiBR2} and~\eqref{eq:babarbelle} with our theoretical predictions. On the experimental side, we observe that the LHCb determination of the branching ratio is more precise but is lower than the B factory average. This may be interpreted as an indication of underestimated higher multiplicity modes in Eq.~\eqref{eq:ExpSemiBR2}. In particular, the uncertainty we assign to these modes in Eq.~\eqref{eq:K3pi} is only 2\% of the inclusive rate. However, comparing the exclusive $K^{(*)}$ branching ratio from LHCb in \eqref{eq:Kavg} and \eqref{eq:Kstaravg} with those from Belle~\cite{Belle:2009zue} and Babar~\cite{BaBar:2012mrf} we also observe a slight tension in the same direction (obscured by the different bin choices). It will be interesting to see whether this tension in the exclusive modes can be resolved with updated (similar bin) measurements from LHCb and Belle~(II) and how the inclusive picture in Fig.~\ref{fig:semicomp} evolves. Moreover, it would be interesting to have measurements of all the $\bar{B}\to K(n\pi)$ exclusive modes used in the sum-over-exclusive determinations with several $q^2$ cuts in order to validate the interpolation between the bins and the Kr\"uger-Sehgal approach.

Finally, comparing Eq.~\eqref{eq:ExpSemiBR2} and Eq.~\eqref{eq:babarbelle} to our theory prediction of the branching ratio in the same region in Eq.~\eqref{eq:branc}, we find good agreement. We note that the uncertainty on the theoretical predictions is over twice as large as the uncertainty given in Eq.~\eqref{eq:ExpSemiBR2}. Comparing to the value extracted using $\cal{R}$ in Eq.~\eqref{eq:TheoryViaR15}, we also find compatibility with the experimental results. Based on the LHCb determination quoted in \cite{Isidori:2023unk} and their $\mathcal{R}$ determination (see discussion in Sec.~\ref{sec:smpred}), the authors of Ref. \cite{Isidori:2023unk} claimed that the tensions in the exclusive $b\to s\ell\ell$ modes at low-$q^2$ are independently verified in the semi-inclusive modes at high-$q^2$, albeit with lower significance. 
However, it is evident that the full picture given by the four determinations in Fig.~\ref{fig:semicomp} does not show any clear tension with the SM. This finding is further illustrated in the next subsection where we compare the different determinations at a common lower cutoff $q_0^2=14.4$ GeV$^2$. 

\subsection{Averaging experimental and theoretical results}
\label{sec:averageexptheo}

In this section, we make a more sophisticated comparison between theory and experiment in the low-$q^2$ and the $q^2 > 14.4$ GeV$^2$ regions. It is reasonable to consider a lower cutoff of $q_0^2 = 14.4$ GeV$^2$, corresponding to a larger hadronic phase space for which the HQE is expected to be under better control.  

\begin{figure}
\begin{center}
\includegraphics[width=\linewidth]{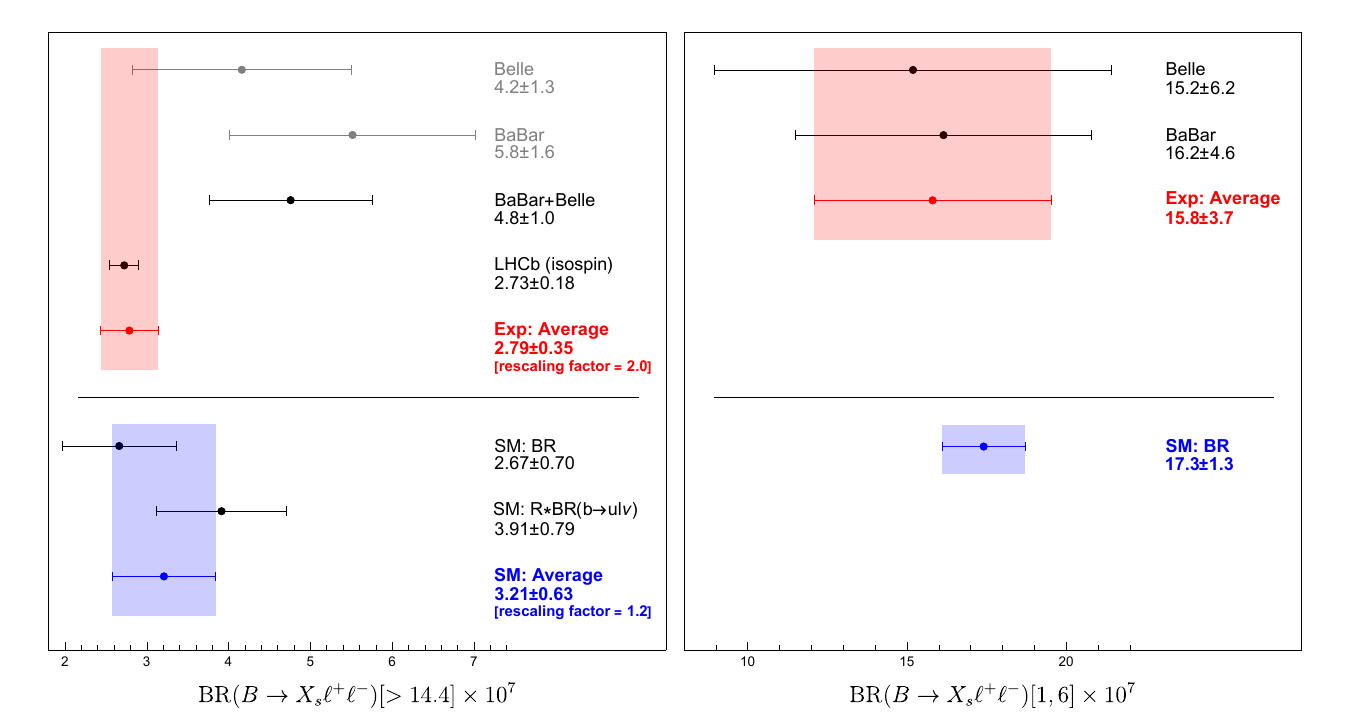}
\caption{ Measurements (red) and SM predictions (blue) for the $\bar B \to X_s \ell^+ \ell^-$ branching ratios in the low-$q^2$ (right) and high-$q^2$ (left) region. Left panel: all results have been adjusted to a common lower endpoint $q_0^2 = 14.4\,\textrm{GeV}^2$ and all results include the logarithmically enhanced QED corrections as described in the text. }
\label{fig:semicomp2}
\end{center}
\end{figure}

The weighted average from BaBar~\cite{BaBar:2004mjt, BaBar:2013qry} and Belle~\cite{Belle:2005fli, Belle:2014owz} in the low-$q^2$  region is 
\begin{equation}
    \mathcal{B}[1,6]_{\rm exp \;average} = (1.58 \pm 0.37) \times 10^{-6} \ .
    \label{eq:brlowexp}
\end{equation}
Here also the average over electron and muon modes is taken.  At low-$q^2$ no equivalent of the sum-over-exclusive measurements in \eqref{eq:ExpSemiBR2} is available from LHCb. For the theoretical prediction, we update our determinations given in Ref.~\cite{Huber:2020vup}, with the updated inputs given in section~\ref{sec:inputs}. As discussed in the introduction  and in more detail in Ref.~\cite{Huber:2020vup}, it is necessary to include the logarithmically enhanced QED corrections for the B-factory setup. We give our updated phenomenological results in Appendix~\ref{sec:app:results}. These results are also summarized in the right panel of Fig.~\ref{fig:semicomp2}, showing excellent agreement between the theoretical predictions and the experimental measurements.

At high-$q^2$, we are dealing with determinations at different cuts and with different setups that do or do not require the inclusion of QED effects. In Fig.~\ref{fig:semicomp2}, we compare all the available measurements rescaled to a common lower endpoint of $q_0^2 = 14.4$ GeV$^2$ and including logarithmically enhanced QED effects. We also average the electron  and muon final states. The rescaling is done using the theoretical expressions for the branching ratios presented in Table~\ref{tab:phenoresults} and Table~\ref{tab:phenoresultsQED}. In a first step we convert the BaBar measurement from $q^2 > 14.2\; {\rm GeV}^2$ (as presented in Ref.~\cite{BaBar:2013qry}) to $q^2 > 14.4 \; {\rm GeV}^2$; the corresponding  conversion factor is 0.958. This way we obtain the effective BaBar measurement quoted in the left panel of Figure~\ref{fig:semicomp2}. We then proceed to average the BaBar and Belle measurements, which now correspond to the same phase space cuts, leading to
\begin{align}
    \mathcal{B}[>14.4]_\textrm{Belle/BaBar} &= (4.8 \pm 1.0) \times 10^{-7} \, .
    \label{eq:BelleBaBar144}
\end{align}

In order to compare all results with the same cut $q_0^2 = 14.4 \; {\rm GeV}^2$ (as done in Fig.~\ref{fig:semicomp2}), we also  need to rescale the determination of this branching ratio obtained from exclusive $B\to (K,K^*)\mu^+\mu^-$ measurement at LHCb for $q^2 > 15\; {\rm GeV}^2$.  Rescaling the original branching fraction given in Eq.~\eqref{eq:ExpSemiBR2}  with the factor given in Eq.~\eqref{eq:eps}, we obtain 
\begin{align}
    \mathcal{B}[>14.4]_\textrm{LHCb (isospin)} &= (2.73 \pm 0.18) \times 10^{-7} \, .
    \label{eq:LHCb144}
\end{align}
Now we are in the position to average the BaBar/Belle determination from~\eqref{eq:BelleBaBar144} and the LHCb (isospin) one from~\eqref{eq:LHCb144} and find 
\begin{equation}
\mathcal{B}[>14.4]_{\rm exp \;average} = (2.79 \pm 0.35) \times 10^{-7} \, . \label{eq:expavg}
\end{equation}
The uncertainty of the weighted average in~\eqref{eq:expavg} has been inflated by a rescaling factor of 2.0. The latter has been obtained following the standard PDG procedure~\cite{10.1093/ptep/ptac097} and is equal to the square root of the reduced chi-square (i.e.\ divided by $N-1$ when averaging $N$ measurements) of the input measurements evaluated with respect to the weighted-average central value (see section 5.2.2 of Ref.~\cite{10.1093/ptep/ptac097}). 

Note that we first average the BaBar and Belle measurements
and then combine the latter with the LHCb (isospin) determination. Alternatively it is possible to combine the three measurements directly (with $N=3$) and obtain $\mathcal{B}[>14.4]=(2.79 \pm 0.26) \times 10^{-7}$. The smaller uncertainty is due to the smaller rescaling factor, 1.5 versus 2.0, obtained while combining the three measurements simultaneously. We prefer the former averaging procedure because it is overall a more conservative approach and it also shows a clear comparison between two different determinations of the inclusive branching ratio, namely the ones reported by the B factories, and the one obtained from exclusive measurements at LHCb.

We compare the result given in Eq.~\eqref{eq:expavg} to our theoretical predictions in Table~\ref{tab:phenoresultsQED}. Averaging the direct branching ratio calculation with that obtained from the ratio $\mathcal{R}$, we find 
\begin{equation}
   \mathcal{B}[>14.4]_{\rm SM \;average} = (3.21 \pm 0.63) \times 10^{-7} \ ,
\end{equation}
which includes an error rescaling factor of $1.2$  and implements the $+0.20$ theory correlation coefficient between $\mathcal{B}[>14.4]$ and $\mathcal{R}(14.4)$ derived from the dependences on parametric inputs listed in Eqs.~\eqref{eq:branc2} and~\eqref{eq:Rzol1}. Figure~\ref{fig:semicomp2} shows a consistent picture of the average of the theory and the average of the experimental  
determinations.


\section{New physics sensitivities}
\label{sec:np}
The sensitivity of $\bar B \to X_s \ell^+ \ell^-$ to physics beyond the Standard Model has improved in several respects since our previous analysis~\cite{Huber:2020vup}. The high-$q^2$ branching ratio in Eq.~\eqref{eq:expavg} now incorporates LHCb measurements of exclusive $B \to K^{(*)}$ modes and is more precise than the Belle/BaBar average in Eq.~\eqref{eq:BelleBaBar144}. In addition, an analysis of the $\bar B \to X_u \ell \bar\nu$ distribution using the full Belle data set~\cite{Belle:2021ymg} has enabled a competitive phenomenological approach to the high-$q^2$ region through the ratio $\mathcal{R}$ defined in Eq.~\eqref{eq:Rzoldef}.

Since $B \to K^{(*)}\ell^+\ell^-$ decay rates into muons relative to electrons~\cite{LHCb:2022vje,LHCb:2022qnv} are consistent with predictions near unity~\cite{Bordone:2016gaq}, we constrain lepton flavor universal coefficients $C_{9,10} = C_{9,10}^\mu = C_{9,10}^e$, parameterized by new physics corrections $C_{9,10}^{\textrm{NP}} = C_{9,10} - C_{9,10}^{\textrm{SM}}$.\footnote{Coefficients are renormalized at the electroweak scale $\mu_0 = 120\,\textrm{GeV}$ in the $\overline{\textrm{MS}}$ scheme.}

To determine the current bounds, only the inclusive $\mathcal{B}[1,6]$, $\mathcal{B}[>14.4]$, $\mathcal{R}(14.4)$ and leptonic $\mathcal{B}(B_s \to \mu \mu)$ constraints are considered. For the experimental inputs, we use the inclusive branching ratios in Eqs.~\eqref{eq:brlowexp},~\eqref{eq:expavg} and~\eqref{eq:BXuexp}, and the average
\begin{align}
\mathcal{B}(B_s \to \mu \mu)_\text{exp average} = (3.45 \pm 0.29) \times 10^{-9}
\end{align}
of the measurements of the leptonic mode~\cite{ATLAS:2018cur, LHCb:2021awg, LHCb:2021vsc, CMS:2022mgd, CDF:2013ezj} from HFLAV~\cite{HFLAV:2022esi}. For the theory, we generated predictions for general $C_{9,10}^\textrm{NP}$ which are consistent with the central values of $\mathcal{B}[1,6]$, $\mathcal{B}[>14.4]$ and $\mathcal{R}(14.4)$ given in Table~\ref{tab:phenoresultsQED} at the Standard Model point ($C_{9,10}^\textrm{NP}=0$). Uncertainties of inclusive observables were assumed to scale with their central values in the $C_{9,10}^\textrm{NP}$ plane. The central value and uncertainty of $\mathcal{B}(B_s \to \mu \mu)$ as functions of $C_{9,10}^\textrm{NP}$ were obtained in Flavio~\cite{Straub:2018kue}, consistent with
\begin{align}
    \mathcal{B}(B_s \to \mu \mu)_\textrm{SM} = (3.66 \pm 0.14)\times 10^{-9}\,  \label{eq:bsmumu}
\end{align}
from Ref.~\cite{Beneke:2019slt} at the point $C_{9,10}^\textrm{NP}=0$.  

\begin{figure}[t]
\centering
\includegraphics[width=0.47\linewidth]{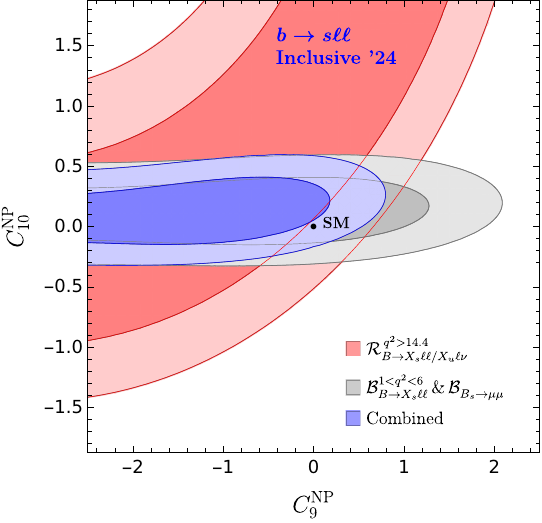} \hspace{0.4cm} \includegraphics[width=0.47\linewidth]{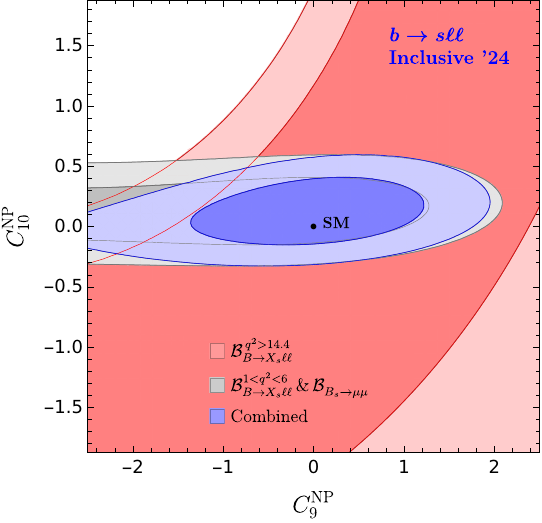}
\caption{Constraints on $b \to s \ell \ell$ coefficients from inclusive and leptonic modes, with (left panel) and without (right panel) normalizing $\mathcal{B}(\bar B \to X_s \ell^+ \ell^-)$ to $\mathcal{B}(\bar B \to X_u \ell \bar\nu)$ in the high-$q^2$ region. Light and dark bands correspond to $68\%$ and $95\%$ confidence intervals. \label{fig:Rfit} }
\end{figure}

\begin{figure}[t]
\centering
\includegraphics[width=0.47\linewidth]{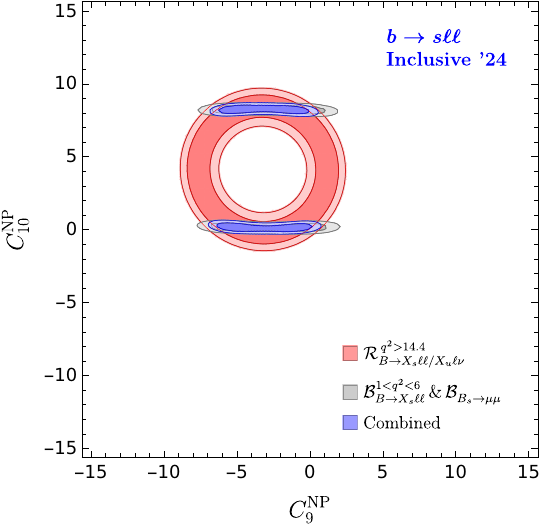} \hspace{0.4cm} \includegraphics[width=0.47\linewidth]{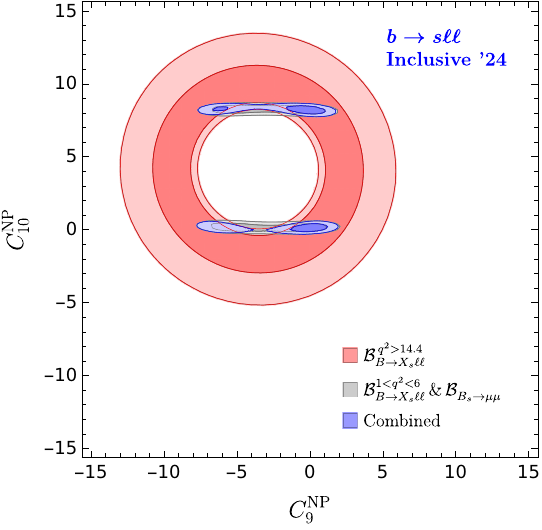}
\caption{Constraints on $b \to s \ell \ell$ coefficients from inclusive and leptonic modes in the expanded plane. See Figure~\ref{fig:Rfit} for further details. \label{fig:RfitLarge} }
\end{figure}

Although $\mathcal{R}(14.4)$ is preferred compared to the branching ratio $\mathcal{B}[>14.4]$ on theoretical grounds, due to the experimental uncertainty on $\bar B \to X_u \ell \bar\nu$, the constraint from the direct determination $\mathcal{B}[>14.4]$ is competitive to the indirect determination. The current constraints on $C_9^{\rm NP}$ and $C_{10}^{\rm NP}$ are shown in Figure~\ref{fig:Rfit}, with (left panel) and without (right panel) using $\bar B \to X_u \ell \bar\nu$ inputs at high-$q^2$. In either approach, both the high-$q^2$ and the combined constraints are consistent with the Standard Model. The same fits in the expanded plane are shown in Figure~\ref{fig:RfitLarge}. The constraint from $\mathcal{B}[>14.4]$ relaxes for unrealistically large new physics deviations since the theory uncertainty is large compared to $\mathcal{R}(14.4)$ and scales with the central value, which becomes large at the boundaries of the panel.

\begin{figure}[htbp]
\centering
\includegraphics[width=0.62\textwidth]{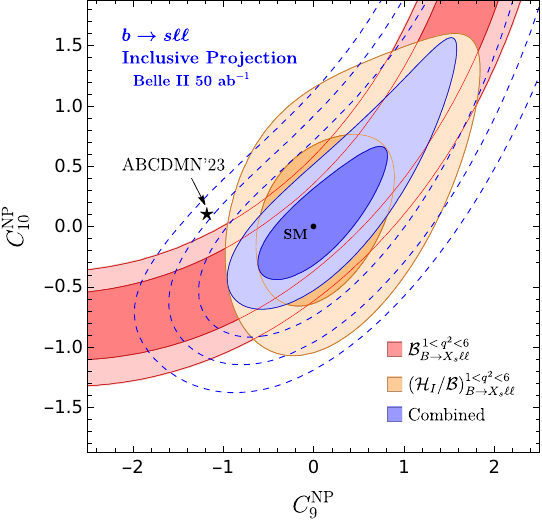} \\
\includegraphics[width=0.62\linewidth]{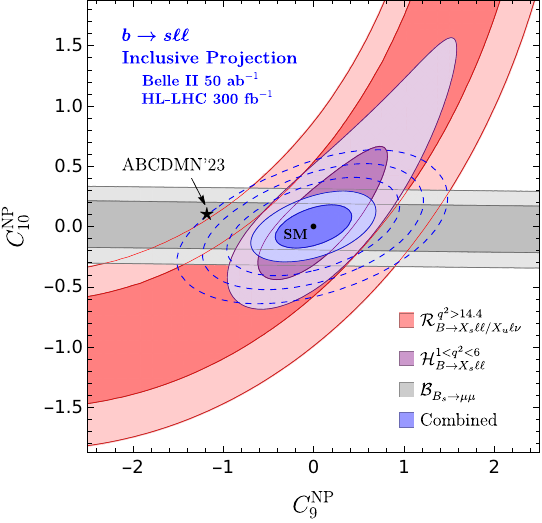}
\caption{Projected constraints on $b \to s \ell \ell$ coefficients centered on the Standard Model. In the lower  panel high-$q^2$ projections at Belle II are combined with the $B_s \to \mu \mu$ projection at the HL-LHC. Light and dark bands of each colour correspond to $68\%$ and $95\%$ confidence intervals; radiating dashed lines are the $3\sigma$, $4\sigma$ and $5\sigma$ contours, respectively. The central value of a fit including exclusive semileptonic modes~\cite{Capdevila:2023hiv} is shown by an asterisk for comparison. See text for further explanations. \label{fig:Belleproj} }
\end{figure}

The decomposition of the inclusive rate into three angular observables will play a crucial role in future precision analyses of $\bar B \to X_s \ell^+ \ell^-$ at Belle II. For the corresponding projection, we adopt the experimental uncertainties for the angular observables in Table~4 of Ref.~\cite{Huber:2020vup} corresponding to $50\,\textrm{ab}^{-1}$ integrated luminosity~\cite{Belle-II:2018jsg}, and assume measurements centered on the predictions in Table~\ref{tab:phenoresultsQED}. In the upper panel of Figure~\ref{fig:Belleproj}, we compare the
projected combined constraint in this region (blue) to the constraints from the rate (red) and angular observables (yellow). We note that in the $C_{10}^\textrm{NP}=0$ scenario, the combined constraint on $C_9^\textrm{NP}$ is significantly stronger than the constraints from the rate and angular observables separately. The angular constraint is a combination of four constraints from two independent normalized angular observables chosen as $\mathcal{H}_T/\mathcal{B}$ and $\mathcal{H}_A/\mathcal{B}$ separated into two bins $1\,\textrm{GeV}^2<q^2<3.5\,\textrm{GeV}^2$ and $3.5\,\textrm{GeV}^2<q^2<6\,\textrm{GeV}^2$. The intersection of the two hyperbolic constraints from $\mathcal{H}_A$ and two circular constraints from $\mathcal{H}_{T}$ (see Fig. 3 of~\cite{Huber:2020vup}) generates the elongated elliptical shape of the yellow band in Figure~\ref{fig:Belleproj} (distorted slightly by the normalization to the rate).  The normalization is conceptually important, as it suppresses the nonperturbative effect of the hadronic mass cut in the low-$q^2$ region~\cite{Huber:2023qse}.

In the high-$q^2$ region, from the current precision in Eq.~\eqref{eq:TheoryViaR14.4} improved by a factor of 3.5 (the expected improvement in inclusive $|V_{ub}|$ determination at Belle II~\cite{Belle-II:2018jsg}), we project $\delta \mathcal{B}(\bar B \to X_u \ell \bar\nu)[>14.4] = 5.2\%$. Combined with $\delta \mathcal{B}(\bar B \to X_s \ell^+ \ell^-)[>14.4] =4.7\%$  in quadrature, one obtains $\delta \mathcal{R}(14.4) = 7.0\%$.\footnote{This marginally updates our projection $\delta \mathcal{R}(14.4) = 7.3\%$ in Ref.~\cite{Huber:2020vup}.} It will be interesting to see whether future determinations of power correction parameters will enable a competitive constraint from $\mathcal{B}[>14.4]$, which is not included in the projection. For the leptonic mode we take $\delta \mathcal{B} (B_s \to \mu\mu) = 4.8\%$ corresponding to $300\,\textrm{fb}^{-1}$ at the HL-LHC~\cite{Cerri:2018ypt}. 

Including the projected constraints from $\mathcal{R}$ and $B_s \to \mu \mu$ (lower panel of Figure~\ref{fig:Belleproj}) contracts the $5\sigma$ contour near the central value of a fit including exclusive semileptonic modes~\cite{Capdevila:2023hiv}. This panel highlights the role of the inclusive mode in scrutinizing whether new short-distance physics is responsible for the anomaly in the exclusive modes. Inclusive measurements with $50\,\textrm{ab}^{-1}$ consistent with the Standard Model would exclude the $C_9^\textrm{NP}\simeq -1$ scenario. However, anomalous inclusive measurements favoring negative $C_9^\textrm{NP}$ would provide compelling evidence in favor of the new physics interpretation of the exclusive anomalies. The current constraints in Fig.~\ref{fig:Rfit} are inconclusive, as they are consistent with the Standard Model as well as scenarios with large negative $C_9^\textrm{NP}$.

Finally, the promising situation at Belle II would benefit from a novel semi-inclusive measurement of $\bar B \to X_s \ell^+ \ell^-$ at LHCb~\cite{Amhis:2021oik}. At least in the high-$q^2$ region, the inclusive mode can be reliably extrapolated from the sum of several exclusive modes, and updated measurements of the exclusive modes can be expected from LHC experiments in the near future. For instance, an analysis of $B^+ \to K^+\ell^+\ell^-$ at CMS has recently become available~\cite{CMS:2024syx}.


\section{Conclusion}
\label{sec:conclusion}

Motivated by a proposed semi-inclusive analysis at LHCb~\cite{Amhis:2021oik}, we have presented predictions for inclusive $\bar B \to X_s \ell^+\ell^-$ observables suitable to the hadron collider environment. It is likely that a semi-inclusive analysis will require simulation of collinear photon radiation, as was the case for exclusive measurements at LHCb, but not at the B factories. To confront the various experimental determinations to each other and to the Standard Model, we have investigated the effect of including collinear photon radiation in our computations, as well as the dependence of the endpoint $q_0^2$ in the high-$q^2$ region $q^2>q_0^2$. In the low-$q^2$ region we also calculate the branching ratio and three angular observables separated into two bins.

The inclusive rate in the high-$q^2$ region is very sensitive to power corrections of order $1/m_b^2$, $1/m_b^3$ and $1/m_c^2$. To reduce the sensitivity to the $1/m_b$ corrections, we calculate the ratio $\cal{R}$ of the $\bar B \to X_s \ell^+\ell^-$ and $\bar B \to X_u \ell \bar\nu$ branching ratios with the same phase space cut~\cite{Ligeti:2007sn}. Multiplying this ratio by the experimental $\bar B \to X_u\ell \bar\nu$ branching ratio results in an indirect determination of the high-$q^2$ branching ratio competitive with the direct theoretical determination, with uncertainty dominated by experimental input rather than power correction parameters. We find that the consistency improves as the endpoint is lowered from $q_0^2=15\,\textrm{GeV}^2$ to $q_0^2=14.4\,\textrm{GeV}^2$, which increases the hadronic phase space, as expected. It will be interesting to see whether the direct and indirect determinations continue to converge with more statistics on the charged current process and more precise determinations of power correction parameters.

We compare our theoretical predictions to a semi-inclusive determination from a sum-over-exclusive $\bar{B}\to K(n\pi)$ modes measured at LHCb. Compared to \cite{Isidori:2023unk}, we also implement the experimental results for higher multiplicity pion final states. We find compatibility between our theoretical determinations of the inclusive rate at high-$q^2$ and the experimental results from the our LHCb determinations and measurements from the B factories. Thus, we do not confirm the claim of the authors of Ref.~\cite{Isidori:2023unk} that the tensions in the exclusive $b \rightarrow s\ell\ell$ modes at low-$q^2$ are independently verified in the semi-inclusive modes at the high-$q^2$, albeit with lower significance.

After correcting for phase space and collinear photon radiation, the average of LHCb and B-factory results significantly improves the sensitivity of the inclusive mode to physics beyond the Standard Model. The resulting constraint from the inclusive and $B_s \to \mu\mu$ modes is consistent with the Standard Model, but new physics corrections to the vector coefficient $C_9$ as large as the Standard Model contribution cannot be excluded yet. We emphasize that the theory uncertainties dominate at high-$q^2$, but in fact originate from sources which are essentially driven by experiment (in particular, $\bar B \to X_u \ell \bar\nu$).

The projected sensitivity to physics beyond the Standard Model with $50\,\textrm{ab}^{-1}$ at Belle II is already quite promising; inclusive measurements consistent with the Standard Model would exclude the $C_9^\textrm{NP} \simeq -1$ scenario at high confidence. This projection would benefit from a dedicated semi-inclusive measurement of $\bar B \to X_s \ell^+ \ell^-$ at the LHC~\cite{Amhis:2021oik}.

\subsubsection*{Acknowledgements}
We thank the organization of the ``Beyond the anomalies'' Workshop in Siegen 2024 which promoted useful discussion with the community about an earlier version of this work. As a consequence, we were able to refine the sum-over-exclusive estimate in \eqref{eq:ExpSemiBR2} in coordination with Gino Isidori, Zach Polonsky and Ariana Tinari. We thank Patrick Owen and Rafael Coutinho for discussions on the experimental LHCb results used for this estimate.
In addition, we thank Patrick Owen, Ulrik Egede and Johannes Albrecht for discussions on the treatment of photons in LHCb. In addition, we thank Florian Bernlochner and Lu Cao for extensive discussions on the treatment of photons in Belle II. We also thank M\'eril Reboud for discussions on statistical analyses. 

The work of QQ was supported in part by the National Natural Science Foundation of China under Grant No.~12375086. The work of EL was supported in part by the U.S.\ Department of Energy under grant number DE-SC0010120. The work of T.~Huber and JJ was supported in part by the Deutsche Forschungsgemeinschaft (DFG, German Research Foundation) under grant  396021762 - TRR 257 ``Particle Physics Phenomenology after the Higgs Discovery''. The work of T.~Hurth was supported by the Cluster of Excellence `Precision Physics, Fundamental Interactions, and Structure of Matter' (PRISMA+ EXC 2118/1) funded by the German Research Foundation (DFG) within the German Excellence Strategy (Project ID 390831469). He also thanks the CERN theory group for its hospitality during his regular visits to CERN where part of this work was written. KKV acknowledges support from the Dutch Research Council (NWO) in the form of the VIDI grant ``Solving Beautiful Puzzles''.

\appendix
\section{Standard Model predictions for the B factories}\label{sec:app:results}

\begin{table}
\begin{center}
$\bar B \to X_s \ell^+\ell^-\;  (\ell=e,\mu\; {\rm average})$\\[0.4em]
\begin{tabular}{|c|c|c|c|}
\hline
\rule{0pt}{12.8pt}
$q^2~\rm{range} \;[{\rm GeV}^2]$ & $[1,6]$ & $[1,3.5]$ & $[3.5,6]$ \\[0.1em]
\hline
\rule{0pt}{12.8pt}
${\cal B}~[10^{-7}]$ & $17.41 \pm 1.31$& $9.58\pm 0.65$& $7.83\pm0.67 $\\[0.3em]
${\cal H}_T~[10^{-7}]$ & $4.77\pm 0.40$& $2.50\pm 0.18$& $2.27\pm 0.22$\\[0.3em]
${\cal H}_L~[10^{-7}]$ & $12.65\pm 0.92$& $7.085\pm 0.48$& $5.56\pm 0.45$\\[0.3em]
${\cal H}_A~[10^{-7}]$ & $-0.10\pm 0.21$ & $-0.989\pm 0.080$ & $0.89\pm 0.16$\\[0.1em]
\hline \hline
\rule{0pt}{12.8pt}
$q^2~\rm{range} \;[{\rm GeV}^2]$ & \multicolumn{3}{c|}{$>14.4$ }\\[0.1em] \hline
${\cal B}~[10^{-7}]$ & \multicolumn{3}{c|}{\rule{0pt}{12.8pt}$2.66\pm 0.70$}\\[0.3em]
$\mathcal{R}(q^2_0)~[10^{-4}]$ & \multicolumn{3}{c|}{$22.27\pm 1.83$ }\\[0.1em] \hline
\end{tabular}
\caption{ Phenomenological results including log-enhanced QED
corrections to the $\bar B \to X_s \ell^+\ell^-$ process. All quantities are obtained by averaging $\ell=e,\mu$. The denominator of the ratio ${\cal R}(q_0^2)$ (i.e.\ the $\bar B \to X_u\ell\bar\nu$ rate for $q^2>q_0^2$), on the other hand, does not include effects which correspond to log-enhanced QED corrections on the theory side. See text for further details.\label{tab:phenoresultsQED} }
\end{center}
\end{table}

In this section, we present updated predictions for several observables including the logarithmically enhanced QED corrections as in~\cite{Huber:2007vv,Huber:2015sra,Huber:2020vup}. We stress again that these predictions are suitable for fully inclusive measurements at the B factories. We update these predictions, which we used in section~\ref{sec:averageexptheo}, in Table~\ref{tab:phenoresultsQED}, in particular due to changes in HQE input parameters for the high-$q^2$ observables. Due to the QED effects, one has to distinguish between the muon and electron final states. Here we average both as is done in the current experimental measurements. 

In addition, the prediction for the ratio $\mathcal{R}$ does not include QED effects for the $b\to u$ transition. This is because we later multiply the measured quantity by \eqref{eq:BXuexp} in which these effects should not be taken into account\footnote{We thank Florian Bernlochner and Lu Cao for discussion about the $b\to u$ measurement.}.


\bibliography{references}{}
\bibliographystyle{JHEP} 
                                                                            
\end{document}